\documentclass[namedreferences]{SolarPhysics}
\usepackage[optionalrh]{spr-sola-addons} 
\usepackage{graphicx} 
\usepackage{color} 
\usepackage{url} 

\newcommand{\etal}{{\it et~al.}}
\newcommand{\eg}{{\it e.g.}}
\newcommand{\ie}{{\it i.e.}}
\newcommand{\etc}{{\it etc.}}
\newcommand{\freq}{\nu}
\def\note #1]{{\bf #1]}}

\renewcommand{\vec}[1]{ {\mathbf #1} }

\newcommand{\ds}{\displaystyle}

\begin{document}

\begin{article}

\begin{opening}

\title{The Current Status of Asteroseismology}

\author{C.~Aerts$^{1}$\sep
 J.~Christensen-Dalsgaard$^{2}$\sep
 M.~Cunha$^{3}$\sep
 D.W.~Kurtz$^{4}$ 
 }
\runningauthor{C.\ Aerts \etal}
\runningtitle{The Current Status of Asteroseismology}

\institute{$^{1}$Instituut voor Sterrenkunde, Katholieke Universiteit Leuven,
	 Celestijnenlaan 200 D, 3001 Leuven, Belgium; Afdeling Sterrenkunde,
	 Radboud University Nijmegen, PO Box 9010, 6500 GL Nijmegen, The
	 Netherlands. email: \url{conny@ster.kuleuven.be}\\
$^{2}$Institut for Fysik og Astronomi, Aarhus Universitet, Aarhus, Denmark.
 email: \url{jcd@phys.au.dk} \\
$^{3}$Centro de Astrof\'\i sica da Universidade do Porto, 
 Rua das Estrelas, 4150-762, Porto, Portugal.
email: \url{mcunha@astro.up.pt}\\
$^{4}$Centre for Astrophysics, University of Central Lancashire, Preston PR1
	 2HE, UK. email: \url{dwkurtz@uclan.ac.uk}
}

\begin{abstract}
Stellar evolution, a fundamental bedrock of modern astrophysics, is driven by
the physical processes in stellar interiors. While we understand these processes
in general terms, we lack some important ingredients. Seemingly small
uncertainties in the input physics of the models, \eg, the opacities or the
amount of mixing and of interior rotation, have large consequences for the
evolution of stars. The goal of asteroseismology is to improve the description
of the interior physics of stars by means of their oscillations, just as global
helioseismology led to a huge step forward in 
our knowledge about the internal structure of the Sun.
In this paper we present the current
status of asteroseismology by considering case studies of stars with a variety
of masses and evolutionary stages. In particular, we outline how the
confrontation between the observed oscillation frequencies and those predicted
by the models allows us to pinpoint limitations of the input physics of current
models and improve them to a level that cannot be reached with
any other current method.
\end{abstract}
\end{opening}

\section{Introduction}
\label{Introduction} 

Despite extensive research in recent decades, we lack detailed knowledge
of some important physical processes relevant for the description of 
stellar interiors. The reason
is that, in general, the existing observations do not yet allow a detailed
confrontation with the description of the physical properties of either the
stellar material in the deepest internal layers, or of the dynamics of the outer
stellar envelope. At first sight, seemingly small uncertainties in the input
physics of the models have large consequences for the whole duration and end of
the stellar life cycle. The lack of a good understanding of interior transport
processes, caused by different phenomena such as rotation, gravitational
settling, radiative levitation, magnetic diffusion, \etc, is particularly
acute when it comes to precise predictions of stellar evolution, and the
galactic chemical enrichment accompanying it.

Given that global helioseismology led to a huge step forward in the accuracy of
the internal structure model and of the transport processes inside the Sun,
asteroseismology aims to obtain similar improvements for different
types of stars by means of their oscillations. Stellar oscillations indeed
offer a unique opportunity to probe the internal properties and processes,
because these affect the observable frequencies. The confrontation
between the frequencies measured with high accuracy and those predicted by
models gives insight into the limitations of the input physics of
models and improves them. In fact, stellar oscillation frequencies are
the best diagnostic known that can reach the required precision in the
derivation of interior stellar properties.

At present, the unknown aspects of the physics and dynamics are dealt with by
using parameterized descriptions, where the parameters are tuned from
observational constraints. These concern, \eg, the treatments of convection, 
the equation of state, diffusion and settling of elements. When a lack
of observational constraints occurs, solar values are often assigned, \eg, to
the mixing length parameter in the description of convection, or phenomena are
ignored, \eg, convective overshooting and the diffusion of heavy elements. It is
hard to imagine, however, that one single 
set of parameters is appropriate for very
different types of stars. Similarly, rotation is either not included or 
is included only
with a simplified treatment of
the evolution of the rotation law in stellar models (\eg, Maeder and
Meynet, 2000). Fortunately, rotation also modifies the frequencies of the star's
modes of oscillation (\eg, Gough, 1981; Saio, 1981). 
An adequate
seismic modelling of rotation inside stars therefore is within reach with
high-precision measurements and mode identifications of stellar oscillations.

The origin and physical nature of stellar oscillations, as well as their
mathematical properties, were thoroughly discussed by Cunha \etal\ (2007), to
which we refer the reader for the theoretical considerations of
asteroseismology. Extensive recent overviews of the occurrence of stellar
oscillations across the entire HR diagram, and asteroseismic applications
thereof, are already available in Kurtz (2004, 2006), Cunha \etal\ (2007), and
Aerts, Christensen-Dalsgaard, and Kurtz (in preparation for Springer-Verlag). 
Rather than repeating
such an observational overview here in a 
far more concise format, we have opted to
outline the current status of asteroseismology by focusing on a few carefully
chosen examples that show the merits this research field has brought to the
improvement of stellar modelling, \ie, we confine ourselves to cases where
quantitative measures of internal structure parameters have been achieved. We
start by considering examples of stars that oscillate similarly to the Sun, and
then move on to pulsators excited by a heat mechanism for a discussion of
convective overshooting and rotation inside stars. The rapidly oscillating Ap
stars, being pulsators with very strong magnetic fields, are of particular
interest to solar astronomers, so are discussed in a separate paper in this
volume (Kurtz, 2008).

\section{From the Sun to Stars: the Properties of Solar-Like Pulsators}
\label{solarlike}

\begin{figure} \begin{center}
\rotatebox{0}{\resizebox{12cm}{!}
{\includegraphics{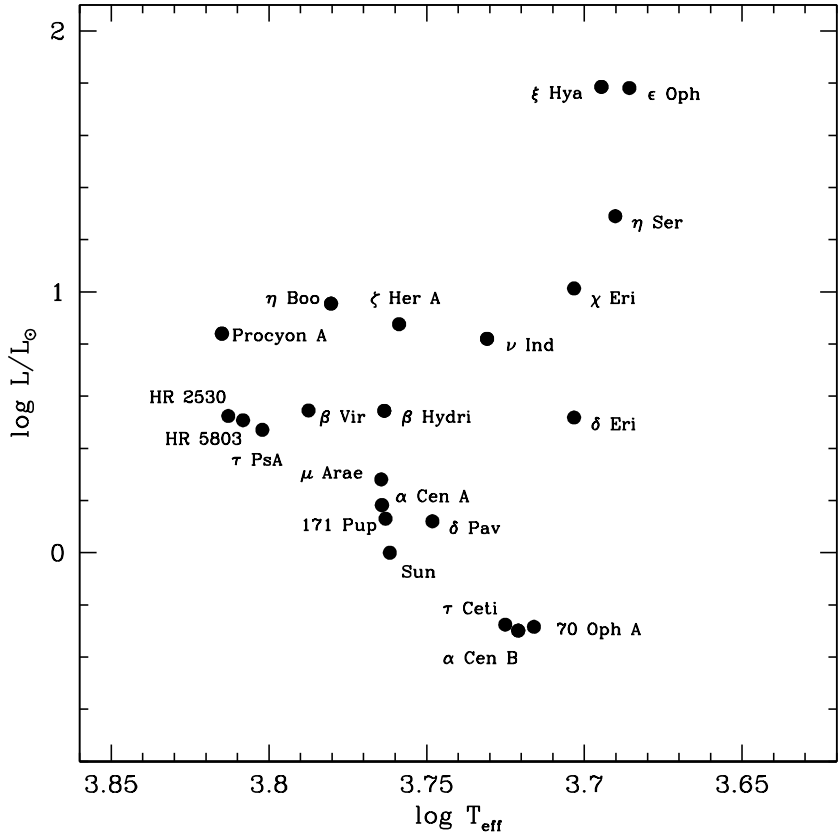}}} \end{center}
\caption[]{HR diagram showing the stars in which solar-like
oscillations have been detected 
{ (with HR\,2530 = HD\,49933). 
The discoveries for 
171\,Pup, HR\,5803 (HD\,139211) 
and $\tau\,$PsA are unpublished (Carrier \etal, in
preparation). }
Figure courtesy of Fabien
Carrier.} \label{fabien1} \end{figure}

As the oscillations of the Sun are caused by turbulent convective motions near
its surface, we expect such oscillations to be excited in all stars with 
significant outer
convection zones. Solar-like oscillations are indeed predicted for the
lowest-mass main-sequence stars up to objects near the cool edge of the
classical instability strip with masses near 1.6\,$M_\odot$ (\eg,
Christensen-Dalsgaard, 1982; Christensen-Dalsgaard and Frandsen, 1983; Houdek 
\etal, 1999) as well as in red giants (Dziembowski \etal, 2001). Such
stochastically excited oscillations have very tiny amplitudes, which makes them
hard to detect, particularly for the low-mass stars. Velocity amplitudes
were predicted to scale roughly as $L/M$,
where $L$ and $M$ are the luminosity and mass of the star,
before the first firm discoveries of
such oscillations in stars other than the Sun (Kjeldsen and Bedding, 1995). This
scaling law was later modified to $(L/M)^{0.8}$ from excitation predictions
based on 3D computations of the outer convection zones 
of the stars (Samadi \etal, 
2005), resulting in lower amplitudes compared with those found for 1D models.

The modes observed in the Sun at low spherical harmonic degree $l$,
and hence solar-like oscillations observable in distant stars, are high-order
acoustic modes.
They satisfy an approximate asymptotic relation
(\eg, Vandakurov, 1967; Tassoul, 1980; Gough, 1993),
according to which, to leading order,
\begin{equation}
\freq_{n\ell} \sim \Delta \freq \left(n + {\ell \over 2} + \epsilon \right) \; ,
\label{asymp}
\end{equation}
where $\freq_{nl}$ is the cyclic frequency of a mode of radial order $n$ and
degree $l$ and $\epsilon$ is a function of frequency determined mainly by
the conditions near the stellar surface.
Also,
\begin{equation}
\Delta \freq = \left( 2 \int_0^R {{\rm d}r \over c} \right)^{-1}
\end{equation}
is a measure of the inverse sound travel time over a stellar diameter,
$r$ being the distance to the stellar centre, $R$ the surface radius of the 
star and $c$ is the adiabatic sound speed.
From simple considerations it follows that 
{\it the large frequency separation\/} satisfies
$\Delta \freq \propto (M / R^3)^{1/2}$
and hence is a measure of the mean density
of the star.
Departures from Equation\,(\ref{asymp}) can be characterized by 
{\it the small frequency separation\/}:
$\delta \freq_{n\ell} = \freq_{n\ell} - \freq_{n-1 \, \ell+2}$.
This quantity is mainly sensitive to the sound speed in the core of the star
and hence provides a measure of the evolutionary state.
The sensitivity of $\Delta \freq$ and $\delta \freq$ on stellar properties
allows a calibration of stellar models in terms of
$(\Delta \freq, \delta \freq)$ to estimate the mass and evolutionary stage
of the star
({\eg}, Christensen-Dalsgaard, 1984, 1988; Ulrich, 1986).

The search for solar-like oscillations in stars in the solar neighbourhood has
been ongoing since the early 1980s. The first indication of stellar power with a
frequency dependence similar to that of the Sun was obtained by Brown \etal\
(1991) in $\alpha\,$CMi (Procyon, F5IV).  The first detection of individual
frequencies of solar-like oscillations was achieved from high-precision
time-resolved spectroscopic measurements only in 1995 for the G5IV star
$\eta\,$Boo (Kjeldsen \etal, 1995); Brown \etal\ (1997), however, could not
establish a confirmation of this detection from independent measurements, but it
was subsequently confirmed by Carrier, Bouchy, and Eggenberger (2003) and
Kjeldsen \etal\ (2003). It took another four years before solar-like
oscillations were definitely established in Procyon (Marti\'c \etal,
1999). 
{ 
While there was a recent controversy about this detection (Matthews \etal,
2004; Bedding \etal, 2005) 
which we do not
discuss in detail here, the results of Marti\'c \etal (1999) have been confirmed
(Mosser \etal, 2008) and an in-depth asteroseismic investigation based on a
large multisite campaign is presently being conducted. Subsequent to Marti\'c et
al.'s results, 
solar-like 
oscillations
} 
were found in two more stars: the G2IV
star $\beta\,$Hyi (Bedding \etal, 2001) and the solar twin $\alpha\,$Cen\,A
(Bouchy and Carrier, 2001). These important discoveries led to several more
subsequent detections, a summary of which was provided by Bedding and Kjeldsen
(2007). The positions of confirmed solar-like pulsators in the HR diagram are
displayed in Figure\,\ref{fabien1}. The detected frequencies and frequency
separations for all stars behave as expected from theoretical predictions and
from scaling relations based on extrapolations from helioseismology.

The oscillation frequencies and frequency separations detected in solar-like
pulsators provide additional constraints with which to test models of stellar
structure and evolution in conditions slightly different from those provided by
the Sun. Such studies generally involve a fit of theoretical models,
characterized by a number of model parameters, to the set of seismic and
non-seismic data available for a given pulsator. Theoretical modelling of
solar-like pulsators using this direct fitting approach has been carried out for
several stars, including $\eta$\,Boo (Carrier, Eggenberger, and Bouchy, 2005;
Guenther \etal, 2005), Procyon (Eggenberger \etal, 2004a; Eggenberger, Carrier,
and Bouchy, 2005; Provost \etal, 2006), and $\alpha$\,Cen\,A and B (\eg,
Eggenberger
\etal, 2004b; Miglio and Montalb\'an, 2005; Yildiz, 2007).  So far, the
main results of these fits are estimates of the stellar masses, ages, and
initial metallicities, even though in some cases the results are still
controversial, and call for better sets of data.

Among the solar-like pulsators, the binary star $\alpha$\,Cen\,A and B provides
a particularly interesting test-bed for studies of stellar structure and
evolution, due to the numerous and precise seismic and non-seismic data that are
available for both components of the binary. Studies of $\alpha$\,Cen\,A and B,
including seismic and non-seismic data for both components, indicate that the
age of the system is likely to be between 5.6 and 7.0\,Gyr, the value derived
being dependent, in particular, on the seismic observables that are included in
the fits.  Moreover, the same studies point to a significant difference in the
values of the mixing-length parameter ($\alpha_{\rm MLT}$, see
Section\,\ref{overshoot} for a definition), for the two stars,
although the sign is uncertain and possibly dependent on the detailed treatment
of the effects of the near-surface layers in the analysis of the observed
frequencies.  Eggenberger \etal\ (2004b) and Miglio and Montalb\'an (2005) found
that the value for $\alpha$\,Cen\,B is larger than that for $\alpha$\,Cen\,A,
whereas Teixeira \etal\ (in preparation) found that $\alpha_{\rm MLT}$ was
slightly {\it smaller\/} for $\alpha$\,Cen\,B than for $\alpha$\,Cen\,A.  The
latter study also found that the best-fitting model for $\alpha$\,Cen\,A was on
the border of having a convective core (see Christensen-Dalsgaard, 2005): even a
slight increase in the mass of the model led to a significant convective core
and hence a model that was quite far from matching the observed properties.

Despite the successful case studies just outlined, the detailed seismic studies
of stars with stochastically-excited modes are currently still in their infancy
compared with global helioseismology. However, given the recent detections and
the continuing efforts to improve them, we expect very substantial progress in
the seismic interpretation of such targets in the coming years. In particular,
the CoRoT (\eg, Michel \etal, 2006) and {\it Kepler\/} 
({\eg}, Christensen-Dalsgaard
{\etal}, 2007) missions will give data of very high quality on solar-like
oscillations.  As seen in the example of $\alpha$\,Cen\,A above, it is
noteworthy that the class of main-sequence solar-like oscillators encompasses
transition objects regarding the development of a convective core on the main
sequence ($1\,M_\odot < M < 1.5\,M_\odot$). Asteroseismology will surely
refine the details of the yet poorly understood physics that occurs near the
core of the objects in this transition region.  Also, data of the expected
quality will provide information about the depth and helium content of the
convective envelope ({\eg} Houdek and Gough 2007a), as well as more reliable
determinations of stellar ages (Houdek and Gough 2007b).  

Additional information from solar-like oscillations is available in the
cases of relatively evolved stars, beyond the stage of central hydrogen burning.
Here the frequency range of stochastically-excited modes may encompass
{\it mixed modes\/}
behaving as standing internal gravity waves, or $g$ modes, in
the deep chemically inhomogeneous regions, thus providing much higher
sensitivity to the properties of this region.
In fact, there is some evidence that such modes have been found in
the subgiant $\eta$\,Boo (Christensen-Dalsgaard, Bedding, and Kjeldsen, 1995).

Evidence for rotational splitting (see Section\,\ref{rotation} for a definition)
has been found in $\alpha$\,Cen\,A (Fletcher {\etal}, 2006; Bazot {\etal},
2007).  However, it has not yet been possible to map the interior rotation of a
solar-like pulsator, since the present frequency multiplet detections are
insufficient.  We note that Lochard, Samadi, and Goupil (2004) found, with
simulated data, that the presence of mixed modes in a star such as $\eta$\,Boo
may allow some information to be derived about the variation of the internal
rotation with position.

For a few pulsators
excited by the heat mechanism, data are already available that provide such
information, albeit only very roughly.  We discuss this further in
Section\,\ref{rotation}, but first we highlight in the next section some case
studies through which the properties of core convection have been tuned by
asteroseismology.

\section{Seismic Derivation of Convective Overshooting inside Stars}
\label{overshoot}

The standard description of convection used in stellar modelling is the Mixing
Length Theory (MLT) of B\"ohm-Vitense (1958). In this theory, the convective
motions are treated as being time-independent.  In the absence of a rigorous
theory of convective motions based on first principles, the convective cells are
assumed to have a mean-free-path length of $\alpha_{\rm MLT}\,H_{p}$, where
$H_{p}$ is the local pressure scale height. The mixing-length parameter depends
on the physics considered in the model and on the specific formulation of the
MLT used. Its value for Model S for the Sun 
of Christensen-Dalsgaard \etal\ (1996) is
$\alpha_{\rm MLT}\simeq 1.99$, using the B\"ohm-Vitense (1958) MLT formulation.

In the context of stellar evolution, it is of crucial importance to quantify the
amount of matter in the fully mixed central region of the star. This amount is
usually derived from the Schwarzschild criterion, which states that convection
occurs in regions where the adiabatic temperature gradient is smaller than the
radiative gradient. However, from a physical point of view, it is highly
unlikely that convective elements stop abruptly at the boundary set by the
Schwarzschild criterion. Rather, their inertia causes them to overshoot
into the adjacent stable area where radiative energy transport takes
place. The amount of such overshooting is, however, largely unknown. For this
reason, it is customary to express it as $\alpha_{\rm ov}\,H_{p}$ where
$\alpha_{\rm ov}$ is expected to be a small fraction of $\alpha_{\rm MLT}$.

The inability to derive a value for $\alpha_{\rm MLT}$ and $\alpha_{\rm ov}$
from a rigorous theoretical description is highly unsatisfactory, particularly
for stars with a convective core, because the total mass of the well-mixed
central region of the star determines its stellar lifetime. This is the reason
why great effort has been, and is being, made to quantify $\alpha_{\rm ov}$,
keeping in mind that we have already a fairly good estimate of $\alpha_{\rm
MLT}$ from the Sun. We describe here the power of asteroseismology to determine
$\alpha_{\rm ov}$.

In the solar case helioseismic analyses have provided constraints on the
overshoot from the solar convective envelope, assuming that this results in a
nearly adiabatic extension of the convection zone followed by an abrupt
transition to the radiative temperature gradient (Zahn, 1991).  Assuming also,
as usual, a spherically symmetric model, such a behaviour introduces a
characteristic pattern in the frequencies in the form of an oscillatory
variation of the frequencies as functions of the mode order.  From the observed
amplitude of this signal, an overshoot region of the nature considered must have
an extent less than around $0.1\,H_p$ (Basu, Antia, and Narasimha, 1994;
Monteiro, Christensen-Dalsgaard, and Thompson, 1994; Christensen-Dalsgaard,
Monteiro, and Thompson, 1995).  It was found by Monteiro,
Christensen-Dalsgaard, and Thompson (2000) that a similar analysis can be
carried out on the basis of just low-degree modes, such as will be observed in
distant stars.

Owing to their sensitivity to the core structure, the low-degree solar-like
oscillations should in principle be sensitive to overshoot from convective
cores.  Models of $\eta$\,Boo without and with overshoot were considered by Di
Mauro \etal\ (2003, 2004).  Although the present observed frequencies are not
sufficiently accurate to provide direct information about the properties of the
core, it was found that for $\alpha_{\rm ov} \ge 0.2$ models could be found in
the central hydrogen-burning stage which matched the observed location in the HR
diagram.  In such models, mixed modes are not expected; thus the definite
identification of mixed modes would constrain the extent of overshoot in the
star.  Straka, Demarque, and Guenther (2005) considered models of Procyon with
various types of core overshoot to determine the extent to which overshoot could
be asteroseismically constrained.

For the $p$-mode diagnostics considered, little sensitivity to overshoot was
found, while the, 
perhaps unlikely, detection of $g$~modes in Procyon, such as have
been claimed in the Sun, would provide much stronger constraints on the
overshoot distance.  As in the case of $\eta$\,Boo, the definite identification
of the star as being on the subgiant branch, {\eg}, from the properties of the
oscillation frequencies, would provide strict constraints on the extent of
overshoot during the central hydrogen-burning phase.  Mazumdar \etal\ (2006a)
made a detailed analysis of the sensitivity of suitable frequency combinations
to the properties of stellar cores and found that the mass of the convective
core, possibly including overshoot, could be determined with substantial
precision, given frequencies with errors that should soon be reached.  Cunha and
Metcalfe (2007) developed diagnostics of small convective cores that may in
principle also provide information about the properties of overshoot; the
detailed sensitivity still needs investigation, however.

\begin{figure} 
\centering
\rotatebox{-90}{\resizebox{6.7cm}{!}{\includegraphics{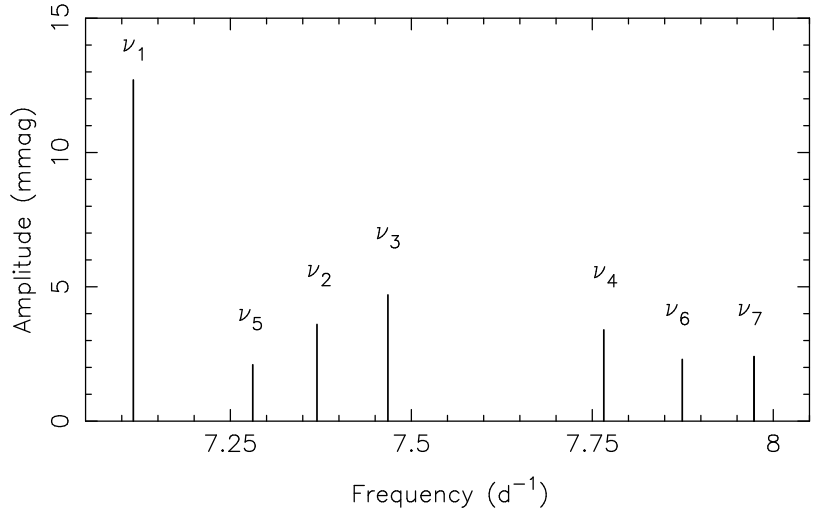}}}
\caption{The schematic frequency spectrum of the $\beta\,$Cep star $\theta\,$Oph
for the Str\"omgren $u$ filter as derived from a multisite photometric
campaign. The measured photometric amplitude ratios led to an identification of
the frequencies $\freq_1, \freq_2, \freq_3$, and $\freq_4$ as, respectively,
$\ell=2,2,0,1$.  (Figure reproduced from Handler, Shobbrook, and Mokgwetsi,
2005).}
\label{thetaoph1}
\end{figure}

\begin{figure} 
\centering
\rotatebox{270}{\resizebox{9cm}{!}{\includegraphics{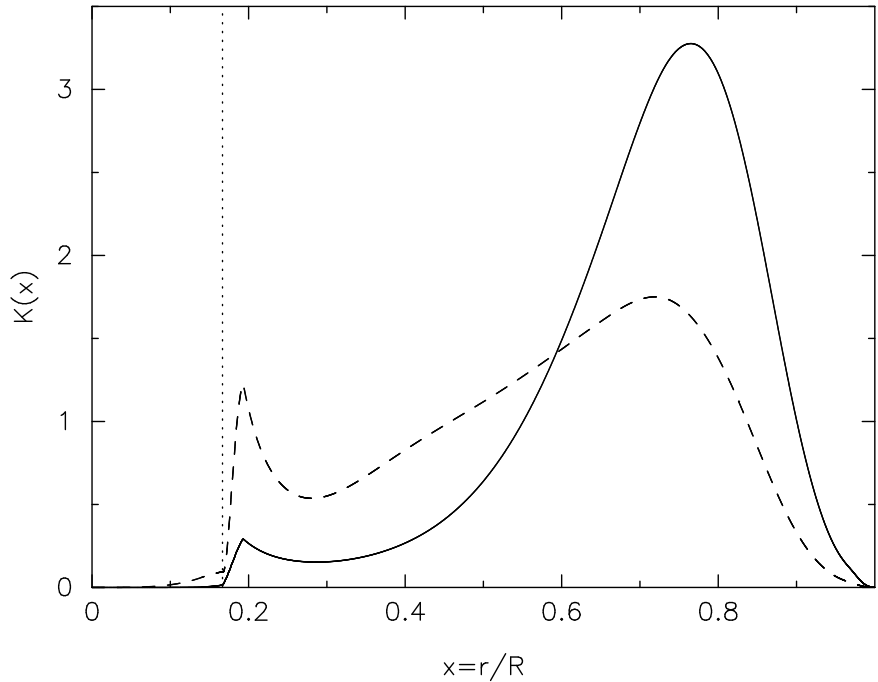}}}
\caption{The rotational kernels defined in Equation\,(\protect\ref{kernel}) as a
function of radial distance inside the star ($x=r/R$), for the identified 
$\ell=1$,
$p_1$ mode (solid line) and $\ell=2$, $g_1$ mode 
(dashed line) of the $\beta\,$Cep
star $\theta\,$Oph. The vertical dotted line marks the position of the boundary
of the convective core, including the overshoot region.  Note that the kernels
also approximately represent the relative sensitivity of the mode frequencies to
other aspects of the stellar interior.  (Figure reproduced 
from Briquet \etal, 2007).  }
\label{thetaoph2}
\end{figure}

Quantitative measures of the core convective overshooting parameter have been
achieved by fitting the frequencies of some of the $\beta\,$Cep stars.  This
group of young, Population~I, near-main-sequence pulsating B stars has been
known for more than a century. They have masses in the range
8\,--\,18\,$M_\odot$, and they oscillate in low-order $p$ and $g$~modes 
with periods
in the range 2\,--\,8 hours. These oscillations are excited by a heat mechanism
acting through opacity features associated with elements of the iron group (\eg,
Dziembowski and Pamiatnykh, 1993; Pamyatnykh, 1999; Miglio, Montalb{\'a}n, and 
Dupret 2007). A
recent overview of the observational properties of the class was provided by
Stankov and Handler (2005). Most of the $\beta\,$Cep stars show multiperiodic
light and line-profile variations and most rotate at only a small fraction of
the critical velocity.

Significant progress in the detailed seismic modelling of the $\beta\,$Cep stars
has occurred over the last few years and has led to quantitative estimates of
the core overshooting parameter $\alpha_{\rm ov}$ for several class members with
slow rotation (see, \eg, Aerts, 2006, for a summary). We illustrate this here
for the star $\theta\,$Oph, whose frequency spectrum was determined from a
multisite photometric campaign and is represented in Figure\,\ref{thetaoph1}
(Handler, Shobbrook, and Mokgwetsi, 2005). An additional long-term,
high-resolution spectroscopic campaign revealed that this star is a member of a
spectroscopic binary with an orbital period of 56.71\,days 
and an eccentricity of
0.17 (Briquet \etal, 2005), and allowed the identification of the spherical
wavenumbers $(\ell,m)$ of the seven detected frequencies from the line-profile
variations induced by the oscillations (see Table\,\ref{thetaophtable}
reproduced from Briquet \etal, 2007).

\begin{table}
\caption{The identification of the pulsation modes of the $\beta\,$Cep star
 $\theta\,$Oph derived from multicolour photometric and high-resolution
 spectroscopic data. Positive $m$-values represent prograde modes.
The amplitudes of the modes are given for the Str\"omgren
 $u$ filter and for the radial velocities. Table reproduced from Briquet \etal\
 (2007). }
\begin{tabular}{ccccrc}
\hline
ID & Frequency (d$^{-1})$ & $(\ell,m)$ & $u$ ampl. & RV ampl. \\
 & & & (mmag) & (km s$^{-1}$)\\
\hline
$\freq_1$ & 7.1160 & $(2,-1)$ & 12.7 & 2.54 \\
$\freq_5$ & 7.2881 & $(2,+1)$ & 2.1 & -- \\
$\freq_2$ & 7.3697 & $(2,+2)$ & 3.6 & -- \\
$\freq_3$ & 7.4677 & $(0,0)$ & 4.7 & 2.08\\
$\freq_4$ & 7.7659 & $(1,-1)$ & 3.4 & -- \\
$\freq_6$ & 7.8742 & $(1,0)$ & 2.3 & -- \\
$\freq_7$ & 7.9734 & $(1,+1)$ & 2.4 & -- \\
\hline
\end{tabular}
\label{thetaophtable}
\end{table}
 
Because the frequency spectra of $\beta\,$Cep stars are so sparse for low-order
$p$ and $g$~modes compared with those of solar-like pulsators (see
Figure\,\ref{thetaoph1}), one does not have many degrees of freedom to fit the
securely-identified modes. This led to the identification of the radial order of
the modes of $\theta\,$Oph as $g_1$ for the frequency quintuplet containing
$\freq_1, \freq_5, \freq_2$, the radial fundamental for $\freq_3$ and $p_1$ for
the triplet $\freq_4, \freq_6, \freq_7$. Fitting the three independent $m=0$
frequencies results in a relation between the metallicity and the
core-overshooting parameter, because the stellar models for main-sequence B
stars typically depend on the five parameters $(X,\ \alpha_{\rm ov},\ Z,\ M,\
{\rm age})$ if we ignore effects of diffusion.  Note that $\alpha_{\rm MLT}$ is
usually fixed to the solar value; for B stars, with their extremely thin and
inefficient outer convection zones, changing $\alpha_{\rm MLT}$ within
reasonable limits does not change the characteristics of the models. In this
way, one finds $\alpha_{\rm ov} = 0.44\pm 0.07$ from a detailed high-precision
abundance determination for $\theta\,$Oph (Briquet \etal, 2007).

The reason why we can derive the core overshooting and the rotation (see
Section\,\ref{rotation}), and provide a quantitative measure of these parameters
for this star, is the different probing ability of the non-radial modes. This
can be illustrated by plotting probing kernels of the modes. Different types of
such kernels are used, depending on the kind of behaviour under
investigation. This is illustrated in Figure\,\ref{thetaoph2}, where we show the
rotational splitting kernels $K(x)$ (which will be defined in
Equation\,(\ref{kernel}) below) of $\theta\,$Oph for the two non-radial
modes. It can be seen that the $g_1$ mode's kernel behaves differently near the
boundary of the core region, and thus probes that region in a different way than
the $p_1$ mode, allowing the derivation of details of the rotational properties
as explained below. A similar figure holds for the energy distribution, which
allows probing the extent of the core region. A comparable result was obtained
for V836\,Cen whose frequency spectrum is almost a copy of that of $\theta\,$Oph
(see Figure\,\ref{hd129929spec}) and also for $\nu\,$Eri (Pamyatnykh, Handler,
and Dziembowski, 2004).

The combination of low-order 
$p$ and $g$~modes thus turns out to be a very powerful
tool to derive the internal structure parameters of massive stars. Additional
measures of the core overshooting have been obtained for the $\beta\,$Cep stars
$\beta\,$CMa (Mazumdar \etal, 2006b) and $\delta\,$Ceti (Aerts \etal, 
2006). For all these $\beta\,$Cep stars, $\alpha_{\rm ov}$ ranges from 0.1 to
$0.5$, although these values depend somewhat on the adopted metal
mixture (Thoul \etal, 2004). It is remarkable that the frequencies of just two
well-identified oscillation modes that have  sufficiently different kernels 
allow one to derive the overshooting parameter with a precision of typically
$0.05$ expressed in $H_{p}$. Adding just a few more well-identified modes should
drastically reduce this error for specific input physics of the
models.

The seismically derived estimates of core overshooting in $\beta\,$Cep stars
are compatible with the
quantitative results for eight detached double-lined eclipsing binaries obtained
by Ribas, Jordi, and Gim{\'e}nez (2000), 
who found $\alpha_{\rm ov}$ to range from 0.1 to
$0.6$ for primary masses ranging from 1.5 to 9\,$M_\odot$. 
Another way of determining the amount of overshooting
from data is by fitting stellar evolutionary tracks to the dereddened
colour-magnitude diagrams of clusters, \eg,
$\alpha_{\rm ov}=0.20\pm 0.05$ for the intermediate age open cluster 
NGC3680 (Kozhurina-Platais \etal, 1997)
and $\alpha_{\rm ov}\approx 0.07$ for the old open cluster 
M67 (VandenBerg and Stetson, 2004). In these
two methods, essentially the same five unknown structure parameters occur as
for the seismic modelling, since
stellar evolution models are used to fit the position of the binary components
and of the cluster main-sequence turn-off point in the HR diagram, respectively.
The uncertainty on the overshoot distance 
derived from the light curve analysis of an
accurately modelled eclipsing binary or from fitting of a cluster turn-off point
is typically between 0.05 and $0.1\,H_{p}$ provided that the metallicities
are known. It is interesting, although perhaps fortuitous,
that all these quantitative
measures of the amount of overshooting are in agreement with the theoretical
predictions by Deupree (2000) from 2D hydrodynamic simulations of zero-age
main-sequence stars with a convective core.

\section{Seismic Derivation of the Internal Rotation Profile of Stars}
\label{rotation}

The rotation of a star implies a splitting of the oscillation frequencies
compared with the case without rotation. Hence, rotation becomes apparent in
frequency spectra as multiplets of $2\ell+1$ components for each mode of degree
$\ell$.  Ignoring rotational effects higher than order one in the rotational
frequency as well as the influence of a magnetic field, the frequency splitting
becomes
\begin{equation}
\ds{\freq_m = \freq_0 + m \int_0^R\ {K}(r)\ {\Omega (r)\over 2 \pi}\ 
\frac{{\rm d}r}{R}} \; ,
\label{split}
\end{equation}
where $\freq_m$ is the cyclic frequency of a mode of azimuthal-order $m$,
and $\Omega$ is the angular velocity 
which we here assume to depend
only on the distance ($r$) to the centre.
The rotational kernels are defined as 
\begin{equation}
\ds{
{K} (r) 
= 
\frac{\left(\xi_r^2-2\xi_r \xi_h +[\ell(\ell+1)-1]\xi_h^2\right) r^2 \rho}
{\ds{\int_0^R \frac{{\rm d}r}{R} 
\left[\xi_r^2 + \ell(\ell+1)\xi_h^2\right] r^2 \rho}}},
\label{kernel}
\end{equation}
with $\xi_r$ and $\xi_h$ the radial and tangential components of the
displacement vector 
\begin{equation}
\vec{\xi} = (\xi_r \vec{e}_r + \xi_h \nabla_h) Y_\ell^m.
\end{equation}

If such
multiplets are observed, their structures are a great help in mode
identification, as non-radial modes with a given value of $\ell$ have 
$2 \ell + 1$ multiplet peaks corresponding to the different values of $m$,
although not all peaks may be visible owing to the geometry
of the modes or excitation of the components,
while radial modes show no multiplet structure. The
recognition of the multiplet structure is far easier for very slow rotators,
where ``slow'' here means that the rotational frequency is far lower than the
frequency spacing for $m=0$ components of modes of adjacent radial order
$n$. 

Below, we describe two types of pulsators for which a quantitative measure of
differential interior rotation has been established.

\subsection{White Dwarfs}

\begin{figure} 
\centering
\rotatebox{-90}{\resizebox{8cm}{!}{\includegraphics{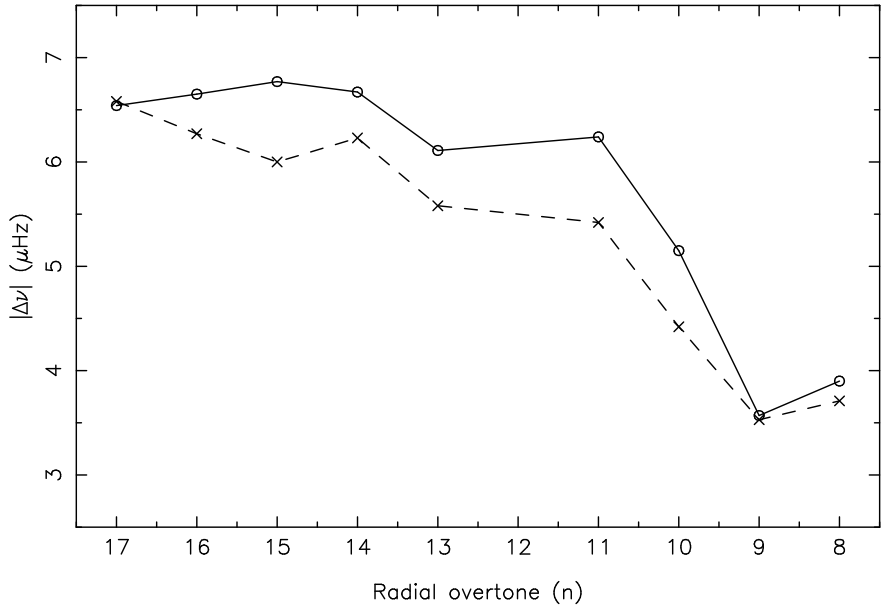}}}
\caption{Frequency splittings $|\Delta \freq| = |\freq_m-\freq_0|$ of the
triplets detected for GD\,358 as a function of radial overtone $n$. The full
line connects the values for the $m=+1$ components corresponding to prograde
modes and the dotted line those for $m=-1$ representing retrograde
modes. (Figure reproduced from Winget \etal, 1994).}
\label{winget}
\end{figure}

The first detection of differential (\ie, non-rigid) rotation inside a star
besides the Sun was achieved for the DBV white dwarf GD\,358 from a multisite
campaign by the Whole Earth Telescope organisation (Winget \etal, 1994). The
multiperiodic variations of DBV white dwarfs are due to low-degree, high-order
$g$~modes, excited by the heat mechanism active in the second partial ionization
zone of helium. Their oscillation periods range from 4 to 12\,minutes and their
photometric amplitudes are relatively large, from a few mmag to 0.2\,mag (\eg,
Bradley, 1995). Among the more than 180 significant frequency peaks detected in
the white-light photometric lightcurve of GD\,358 covering 154\,hours of data,
27 are the components of well-identified triplets. These frequency splittings
are shown as a function of radial order $n$ in Figure\,\ref{winget}. It can be
seen that larger splittings occur for higher radial order, while one would
expect these splittings to be constant for rigid rotation inside the white
dwarf. Since the modes of higher radial-order probe predominantly the outer
layers and those of lower radial-order the inner parts, the rotation of GD\,358
must be radially differential.  The mean splitting for the modes of $n=16$ and
17 leads to a rotation period of 0.89\,days through Equations\,(\ref{split}) and
(\ref{kernel}), while for $n=8,9$ the rotation period is 1.6\,days. Winget
\etal\ (1994) therefore concluded that the inner parts of GD\,358 rotate 0.6
times more slowly than its outer layers, where ``inner'' and ``outer'' refer to
those regions probed by the detected triplets.  However, Kawaler, Sekii, and
Gough (1999) found that the data were not yet of sufficient quality to allow a
more detailed inversion for the variation of the internal rotation with depth.

The detailed seismic modelling of GD\,358 followed that achieved previously
for the prototypical DOV white dwarf PG1159-035 (GW\,Vir) described in the
seminal work by Winget \etal\ (1991), which was again based on data collected
by the Whole Earth Telescope consortium (Nather \etal, 1990). This led to 125
significant frequencies for GW\,Vir, of which 101 were identified as 
components of rotationally split triplets and quintuplets. 
Unlike the case for GD\,358, all multiplets for a given $\ell$
showed the same frequency splitting within the measurement errors, allowing 
Winget \etal\ (1991) to deduce a constant rotation period of 
approximately $1.38\pm0.01$\,day
throughout the white dwarf. Classical spectroscopy can in no way reveal the
rotation periods of single compact stellar remnants with such high precision,
not even in the case where the inclination angle can be estimated from
independent information.

The deviation of GD\,358's splittings for $m=+1$ with respect to those for
$m=-1$ in Figure\,\ref{winget} was interpreted by Winget \etal\ (1994) in terms
of a weak magnetic field of $1300\pm 300$\,G, which causes splittings $\sim
|m^2|$ in addition to the rotational splitting given in Equation\,(\ref{split}) 
(\eg, Dziembowski and Goode, 1984; Jones \etal, 1989). The effect of a magnetic 
field could not be established 
for the
frequency multiplets of PG\,1159-035, which led to an upper limit of 6000\,G for
that object's magnetic field (Winget \etal, 1991). It is noteworthy that the
magnetic-field strength that can be probed by classical spectroscopy of white
dwarfs through the Zeeman effect requires fields roughly a factor of 
1000 stronger
than what can be found from asteroseismology.

The case studies of GD\,358 and PG\,1159-035 by the Whole Earth Telescope
consortium implied not only a first test case for the technique of
asteroseismology, but at the same time a real breakthrough in the derivation of
white dwarf structure models. It not only led to estimates of internal rotation
and magnetic field strength, but also allowed a high-precision mass estimate
($0.586\pm0.003$\,$M_\odot$ for PG1159-035 and $0.61\pm0.03$\,$M_\odot$ for
GD\,358). It also proved that the outer layers of white dwarfs are
compositionally stratified. This was derived from deviations of the frequency
spacings due to mode trapping compared with spacings for unstratified models.
Mass estimates with such high precision cannot be achieved from other means,
except for relativistic effects in binary pulsars. These two seismic
studies of white dwarfs paved the road for many others of their kind, but none
of the more recent ones have led to more accurate internal rotation rates than
those for PG1159-035 and GD\,358. We refer to Kepler (2007) and Fontaine and
Brassard (in preparation) for recent review papers on white-dwarf seismology.

\subsection{Main-Sequence Stars}

There are presently only three main-sequence stars, besides the Sun, for which
an observational constraint on the internal-rotation profile has been
derived. In all three cases, it was achieved through asteroseismology of
$\beta\,$Cep stars. Several of these are suitable targets to attempt mapping of
their interior rotation because their rotational frequencies are well below the
frequency spacing between multiplets. These stars are particularly interesting
targets for this purpose, because the largest uncertainty in stellar evolution
models for massive stars is precisely concerned with rotational mixing effects.

\begin{figure} 
\centering
\rotatebox{-90}{\resizebox{8.cm}{!}{\includegraphics{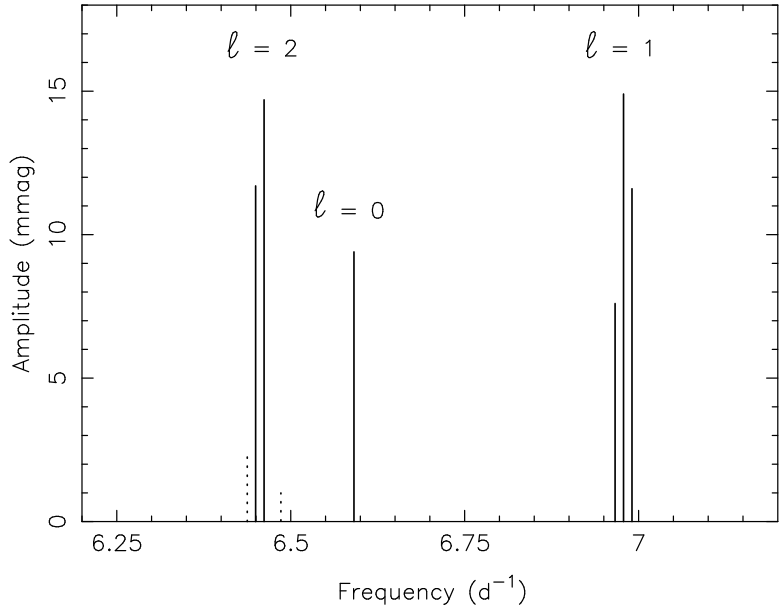}}}
\caption{The schematic frequency spectrum of the $\beta\,$Cep star V836\,Cen
derived from single-site Geneva $U$ data spanning 21\,years. 
The dotted lines are
frequencies that are not yet firmly established; these were not used in the
seismic modelling. (Figure reproduced from Aerts \etal, 2004).}
\label{hd129929spec}
\end{figure}

The first seismic proof of differential rotation in a massive star was obtained
for the B3V star V836\,Cen (HD\,129929; Aerts \etal, 2003). This result was
derived from the well identified (parts) of one rotationally split frequency
triplet and one quintuplet, as shown in Figure\,\ref{hd129929spec}. This star's
detected frequency spectrum is obviously very similar to the one of
$\theta\,$Oph (compare Figures\,\ref{thetaoph1} and \ref{hd129929spec}), except
that V836\,Cen is a slower rotator than $\theta\,$Oph. Given that only two
multiplets were available for V836\,Cen, Dupret \etal\ (2004) assumed a linear
rotation law and concluded that the rotational frequency near the stellar core
is 3.6 times higher than at the surface. It was possible to derive this because
the kernels of the $g_1$ and $p_1$ modes probe differently the rotational
behaviour near the stellar core, just as for $\theta\,$Oph (see
Figure\,\ref{thetaoph2}).  A very similar result, the rotation of the deep
interior exceeding the surface rotation by a factor between three and five, was
obtained by Pamyatnykh, Handler, and Dziembowski (2004) from the $g_1$ and $p_1$
$\ell=1$ modes of the B2III $\beta\,$Cep star $\nu\,$Eri (HD\,29248).  These
results are compatible with the assumption of local angular-momentum
conservation. Both V836\,Cen and $\nu\,$Eri are -- for upper main-sequence stars
-- very slow rotators, with surface rotation velocities of 2\,km\,s$^{-1}$
(V836\,Cen) and 6\,km\,s$^{-1}$ ($\nu\,$Eri). This made the seismic derivation
of the interior rotation possible, because the splitting of the multiplets does
not interfere with the frequency separation between different multiplets.  The
rotation profile itself could not be tuned further, given that only parts of
very few multiplets were available. Classical spectroscopy, even at extremely
high resolution, could never have led to the proof of differential interior
rotation, as it can only measure the surface rotation. Moreover, the intrinsic
line broadening of such stars is typically of order $\approx 10\,$km\,s$^{-1}$,
which is larger than the surface rotation velocity of these two stars,
preventing a derivation of the projected equatorial rotation velocity to better
than 1\,km\,s$^{-1}$.

For the star $\theta\,$Oph, which is a twin of V836\,Cen as far as the detected
frequency spectrum is concerned, rigid interior rotation could not be excluded
from comparison of the frequency spacing in its triplet and its quintuplet
(Briquet \etal, 2007). Its
frequency precision is two orders of magnitude lower that for V836\,Cen and one
order of magnitude lower than for $\nu\,$Eri. In any case, strong differential
rotation is excluded for that star as well.

\section{Expected Future Improvements}

\subsection{Compact Pulsators and the Tuning of Atomic Diffusion}

{ 
A field within asteroseismology, which we did not discuss extensively here
but which is undergoing rapid growth and may tune our knowledge of microscopic
diffusion for stellar structure and of binary star evolution, is the application
to pulsating subdwarf B stars (sdBVs). While sdBVs with $p$~modes were
discovered a decade ago (Kilkenny \etal, 1997), those with $g$~modes were
discovered more recently (Green \etal, 2003). The existence of sdBVs was
predicted independently and simultaneously with their observational discovery
(Charpinet \etal, 1996). An opacity bump associated primarily with iron-group
elements turns out to be
turns out
to be an efficient driving mechanism. The 
atomic diffusion
processes that are at work in sdB stars -- radiative levitation
and gravitational settling -- cause iron (and also zinc) to become overabundant
in the driving zone, thus exciting low-order $p$ and $g$~modes 
(Charpinet \etal, 
1997; Jeffery and Saio, 2006). The details of the diffusion processes are,
however, still uncertain. These may also be relevant for the SPBs and
$\beta\,$Cep stars (Bourge \etal, 2006), for which diffusion processes have
been ignored so far in the seismology. Such processes are dominant in the 
atmospheres of the roAp stars, as is discussed by Kurtz (2008). 

From an evolutionary point of view, the sdB stars are poorly understood. Their
effective temperatures are in the range 23\,000\,--\,32\,000\,K, 
and their $\log
g$ in the range 5\,--\,6. They all have masses below 0.5\,$M_{\odot}$ which
implies that
they have lost almost their entire hydrogen envelope at the tip of the
red-giant branch.  Their thin hydrogen layer does not contain enough mass to
burn hydrogen, making them evolve immediately from the giant branch towards the
extreme horizontal branch. While it is clear they will end their lives as
low-mass white dwarfs, it is yet unclear how they expelled their envelopes. All
scenarios that have been proposed involve close binary interaction (Han \etal,
2003; Hu \etal, 2007).

The currently known sdB pulsators have multiple periods 
in the range 80\,--\,600 seconds 
and amplitudes up to 0.3\,mag. Their amplitude variability and faintness
have prevented unambiguous mode identifications so far, limiting the power of
seismic inference to tune the diffusive and rotational processes. Rapidly
rotating cores have been claimed for some of the sdBVs in order to explain their
dense frequency spectra in terms of low-degree modes (Kawaler and Hostler, 
2005). Firm observational proof of that is not yet available, but, given the
impressive efforts undertaken 
to understand the internal and atmospheric structure of
these stars as well as their evolutionary status, we expect rapid progress in
the near future. For a recent overview of the status of sdB seismology, we refer
to Charpinet \etal\ (2007).

Recently, a seven-year study of the sdBV star V391\,Peg (Silvotti \etal, 2007)
used the extreme frequency stability of two independent pulsation frequencies to
show the presence of a $\approx$3.2-M$_{\rm Jupiter}$ planet that had moved from
about 1\,AU out to 1.7\,AU during the red giant phase of the sdB star precursor,
allowing the planet to survive, much as the Earth may survive the Sun's red
giant phase in about 7\,Gyr.  This novel application of asteroseismology
highlights the close relation and mutual interests of helioseismology,
asteroseismology, planet-finding and solar system studies.

\subsection{Heat-Driven Pulsators along the Main Sequence}

Overshooting parameters and internal rotation profiles have not yet been
determined for the other heat-driven pulsators known along the main sequence
(besides the $\beta\,$Cep stars), such as the slowly pulsating B stars (SPBs)
and the A- and F-type $\delta\,$Sct and $\gamma\,$Dor stars. The main obstacles
to overcome are the limited number of detected oscillation frequencies of the
$g$~modes for the SPB stars and $\gamma\,$Dor stars, and reliable mode
identification for those as well as for the $\delta\,$Sct stars.  While the
pioneering space missions WIRE (Buzasi, 2002; Bruntt and Southworth, 2008) 
and MOST (Matthews \etal, 2004; Walker, 
2008) led to an impressive and unprecedented number of oscillation modes for
several such stars, the time base of the data was limited to a few weeks and
unique mode identifications are not available for these mission's target stars.
It is to be expected that the uninterrupted photometry obtained by the CoRoT
(five months time base, launched 26 December 2006) and {\it Kepler\/} (3.5 years
time base, to be launched in 2009) space missions, along with their ground-based
spectroscopy programmes, will result in the necessary frequency precision and
mode identification.  This should imply big steps forward for the seismic
modelling of these type of stars.

Thus, even with several space missions in operation, ground-based efforts to
increase the number of heat-driven pulsators with (preferably simultaneous)
long-term multicolour photometric and high-resolution spectroscopic data for
mode identification should definitely be intensified. It was this type of
extensive data that yielded sudden and immense progress in the $\beta\,$Cep star
seismology discussed in this paper and that also advanced significantly the
interpretation of the oscillation spectrum of the prototypical $\delta\,$Sct
star FG\,Vir (Zima \etal, 2006).  Only systematic and dedicated observing
programmes can bring us to the stage of mapping and calibrating the
internal-mixing processes inside stars across the HR diagram.

\subsection{Solar-Like Pulsators}

A new dimension in the progress for stochastically-excited pulsators is
expected from the combination of asteroseismic and interferometric data, as
explained by Cunha \etal\ (2007).  The CoRoT and {\it Kepler\/} missions will
also provide a large improvement in the data for solar-like oscillators.  In
particular, {\it Kepler\/} will yield data over several years for more than a
hundred stars, together with three-month surveys of many more stars.  The very
extended observations may reveal possible frequency variations associated with
stellar magnetic cycles, as has been observed in the Sun, and hence improve our
understanding of such cycles.  Also, the identification and interpretation of
mixed modes, which have both a $p$-mode and a $g$-mode character due to a highly
condensed stellar core, would be a great help to tune stellar evolution models
towards the end of, and after, the central hydrogen-burning phase.

Data of even higher quality on
solar-like oscillators can be obtained with
dedicated m\,s$^{-1}$-precision radial-velocity campaigns,
since the intrinsic stellar noise background is much lower, relative to
the oscillations, in velocity than in photometry (e.g.\ Harvey 1988).
This is the goal of the 
SIAMOIS (Mosser \etal, 2007) and
SONG (Grundahl \etal, 2007) projects.  
SIAMOIS will operate from the South Pole, while 
SONG aims at establishing a global network of moderate-sized telescopes. Both
are 
dedicated to obtain high-precision radial-velocity observations.
This will increase substantially the number of (rotationally
split) detected modes, particularly at relatively low frequency where
the mode lifetime is longer and the potential frequency accuracy is higher.
With the high-quality data expected from CoRoT, {\it Kepler}, SIAMOIS  
and SONG, we may hope to
carry out inverse analyses (e.g., Basu \etal, 2002, Roxburgh and Vorontsov 2002)
to infer the detailed properties of stellar cores.

These projects, and others further into the future, covering stars across the
HR diagram will provide an extensive observational basis for investigating
stellar interiors.
Together with the parallel development of stellar modelling techniques we
may finally approach the point, in the words of Eddington (1926), of being
`competent to understand so simple a thing as a star'.

}

\begin{acks}
CA is supported by the Research Council of the Catholic University of Leuven
under grant GOA/2003/04.  MC is supported by the EC's FP6, FCT, and FEDER
(POCI2010) and through the project POCI~/CTE-AST~/57610~/2004. The authors
acknowledge support from the FP6 Coordination Action HELAS.
\end{acks}

{}
\end{article} 
\end{document}